\documentclass[]{spie}  

\usepackage[section]{placeins}
\usepackage{amsmath,amsfonts,amssymb}
\usepackage{graphicx}
\usepackage{booktabs}
\usepackage[colorlinks=true, allcolors=blue]{hyperref}

\title{WST instrument Exposure Time Calculator: full simulation of multi-mode
spectrograph performance from source to detector}

\author[a]{Matteo Ferro}
\author[a]{Matteo Genoni}
\author[b]{Henri M.\ J. Boffin}
\author[c]{Jose Schiappacasse-Ulloa}
\author[c]{Sofia Randich}
\author[d]{Roland Bacon}
\author[b]{Vincenzo Mainieri}
\author[e,f]{Alessio Mucciarelli}
\author[e,f]{Carmela Lardo}
\author[e,f]{Michele Moresco}
\author[e,f]{Cristian Vignali}
\author[e,f]{Margherita Talia}
\author[a]{Andrea Scaudo}
\author[a]{Marco Landoni}
\author[g]{Roelof S.\ de Jong}
\author[g]{Olga Bellido-Tirado}

\affil[a]{INAF -- Osservatorio Astronomico di Brera, Via E.\ Bianchi 46, 23807 Merate (LC), Italy}
\affil[b]{European Southern Observatory, Karl-Schwarzschild-Str.\ 2, 85748 Garching, Germany}
\affil[c]{INAF -- Osservatorio Astrofisico di Arcetri, Largo E.\ Fermi 5, 50125 Firenze, Italy}
\affil[d]{Centre de Recherche Astrophysique de Lyon, Université de Lyon, 9 avenue Charles André, 69230 Saint-Genis-Laval, France}
\affil[e]{Dipartimento di Fisica e Astronomia ``Augusto Righi'', Università di Bologna, Via Gobetti 93/2, 40129 Bologna, Italy}
\affil[f]{INAF -- Osservatorio di Astrofisica e Scienza dello Spazio di Bologna, via Piero Gobetti 93/3, 40129 Bologna, Italy}
\affil[g]{Leibniz-Institut für Astrophysik Potsdam (AIP), An der Sternwarte 16, 14482 Potsdam, Germany}

\authorinfo{Send correspondence to M.\ Ferro: matteo.ferro@inaf.it}

\begin{document} 
\maketitle

\begin{abstract}
We present a comprehensive Exposure Time Calculator (ETC) developed for the
Wide-field Spectroscopic Telescope (WST) concept.  The WST, currently in its
conceptual phase, is designed as a next-generation large spectroscopic survey
facility featuring three complementary observing modes: an Integral Field
Spectrograph (IFS) covering 370--930\,nm at $R\!\sim\!4800$; a high-resolution
Multi-Object Spectrograph (MOS-HR) with four bands at $R\!\sim\!40{,}000$; and a
low-resolution Multi-Object Spectrograph (MOS-LR) with four channels at
$R\!\sim\!3800$--$4900$.
The ETC simulates the complete photon-propagation path from astronomical source
to detector, incorporating wavelength-dependent system throughput (telescope
transmission, instrumental optics, detector quantum efficiency), accurate sky
background via ESO SkyCalc\,\citenum{Noll12,Jones13} integration, and
a comprehensive noise treatment (photon noise, sky background, read-out noise,
dark current).  The computational core is implemented as the {\ttfamily
pyetc\_wst} Python library built on the MPDAF
framework\,\citenum{Bacon16_mpdaf}, supporting multiple target spectral energy
distributions (stellar templates, blackbody, power-law, emission lines, and
user-uploaded spectra with arbitrary redshift) and spatial morphologies (point
sources and S\'ersic extended profiles).
Four operational modes enable flexible exposure-time optimization.  Full spectral
outputs include wavelength-dependent signal-to-noise ratio (SNR), source and sky photon counts, noise
decomposition by component, and simulated extracted spectra.  An interactive web
interface, together with a REST API and a command-line tool, complete the user
experience and enable batch survey-design workflows.
\end{abstract}


\section{INTRODUCTION}
\label{sec:intro}

The Wide-field Spectroscopic Telescope (WST) is a proposed next-generation
facility designed to conduct massively multiplexed spectroscopic surveys across
the full optical domain.  The WST concept\,\citenum{WST2024}
foresees a 12\,m-class telescope equipped with three simultaneously operated
spectrograph systems: an Integral Field Spectrograph (IFS), a low-resolution
Multi-Object Spectrograph (MOS-LR), and a high-resolution Multi-Object
Spectrograph (MOS-HR).  The science case spans stellar astrophysics, galaxy assembly and evolution, large-scale structure, and transient follow-up.

Accurate prediction of the achievable signal-to-noise ratio (SNR) for a given
observing configuration is an essential tool at every phase of an astronomical facility project, from initial concept studies and survey strategy optimization to
individual program preparation. An Exposure Time Calculator must faithfully
reproduce all photon-budget contributions: source emission, atmospheric
transmission and sky radiance, instrument throughput, and detector characteristics.

We present {\ttfamily pyetc\_wst}, a Python package that implements a full
end-to-end ETC for all WST modes.  Section~\ref{sec:instrument} summarizes the
WST instrument suite.  The computational model is described in
Section~\ref{sec:model}.  The software architecture, web interface, and
programmatic access are discussed in Section~\ref{sec:software}.
Example performance predictions are shown in Section~\ref{sec:results}, and conclusions in Section~\ref{sec:conclusions}.

\section{THE WST FACILITY SUITE}
\label{sec:instrument}
\begin{figure}[ht]
\begin{center}
\includegraphics[width=0.66\textwidth]{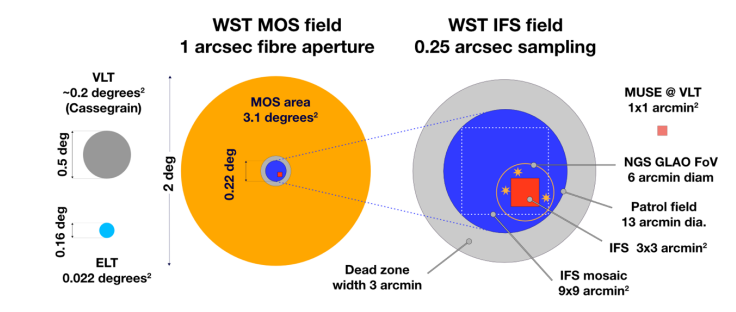}
\end{center}
\caption{\label{fig:MOS_IFS_label}
WST fields of view compared to other facilities. The MOS covers a 2-degree diameter focal plane (3.1 deg$^2$), with a central zone reserved for the IFS. The IFS science field is 3×3 arcmin$^2$ with 0.25 arcsec spaxel sampling, movable within a 13 arcmin patrol field to enable 9×9 arcmin$^2$ mosaic mode. A dead zone of 3 arcmin separated the IFS and the MOS. The GLAO natural guide star annulus (6 arcmin diameter) surrounds the science field. For reference, the VLT/MUSE 1×1 arcmin$^2$ and ELT fields are shown.}
\end{figure}

WST has undergone an intense phase of conceptual and trade-off studies, and the facility design briefly presented here is the outcome of such studies, while details on the FoV in the different instrument modes are given in Figure~\ref{fig:MOS_IFS_label}. The three instruments — IFS, MOS-LR, and MOS-HR — are designed to operate simultaneously. Their key parameters, as implemented in {\ttfamily pyetc\_wst}, are summarized in Table~\ref{tab:instruments}. In the following subsections, a brief description of the telescope and each instrument is provided.

\subsection{Telescope}
\label{sec:telescope}
\begin{figure}[ht]
\begin{center}
\includegraphics[width=0.8\textwidth]{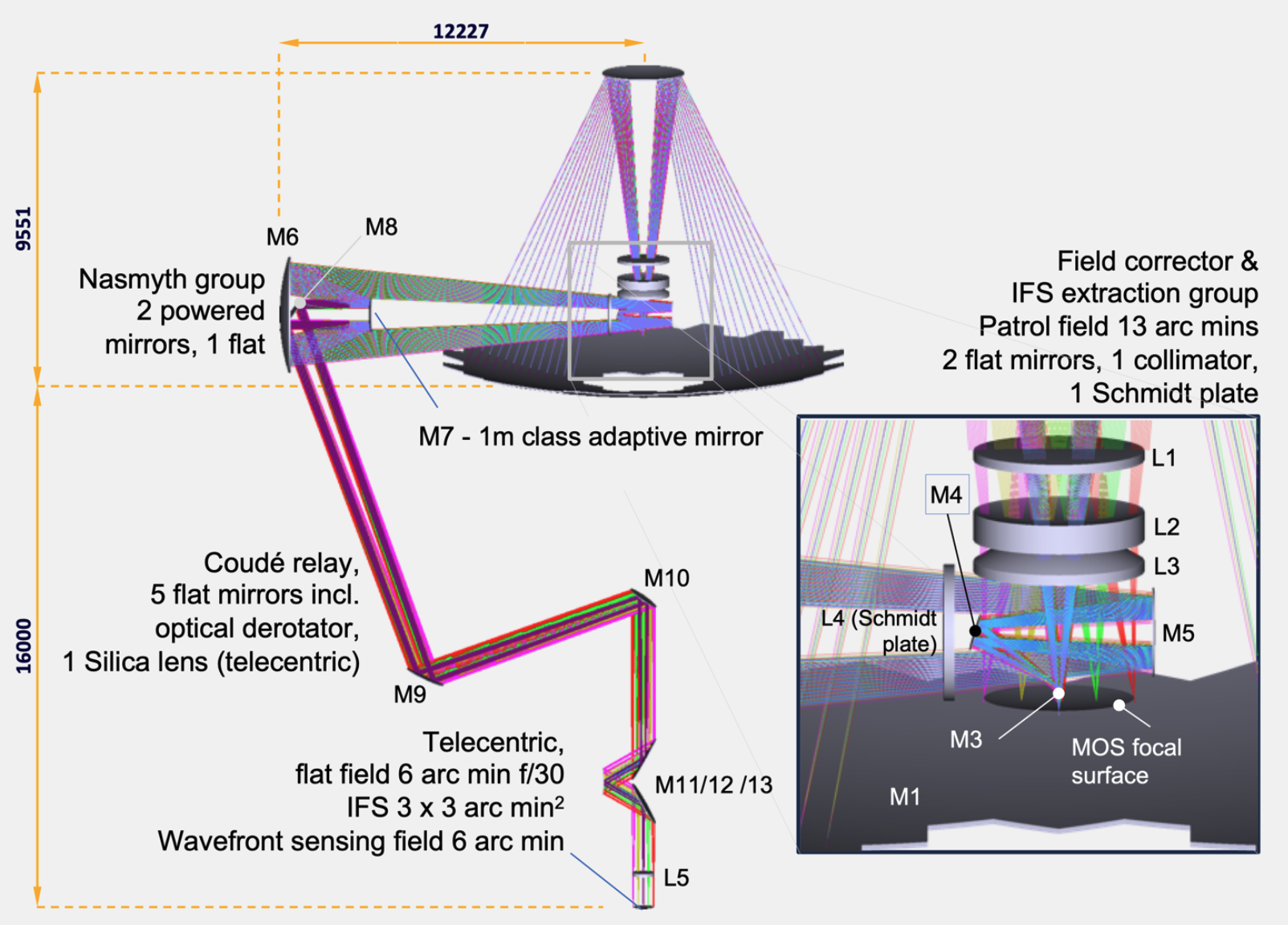}
\end{center}
\caption{\label{fig:des}
Reference telescope optical design.}
\end{figure}

The telescope primary mirror is a 12-m segmented mirror (78 segments of 1.4\,m, closely based on the ELT segment design) working at f/1.0 in Cassegrain configuration. A key aspect of the design is the delivery of two simultaneous focal planes: a MOS focal plane providing a large 2-degree field of view, and a central IFS focal plane of 6\,arcmin diameter, encompassing the $3\times3$\,arcmin$^2$ science field together with the surrounding 6-arcmin circular region used to acquire natural guide stars for the Ground-Layer Adaptive Optics (GLAO) system. The IFS science field can be located anywhere within a 13\,arcmin diameter patrol field. The full telescope optical path is shown in Figure~\ref{fig:des}, and a detailed description is given in \citenum{WST2026-Telescope} and \citenum{WST2026-Roland}.

The MOS focal plane delivers an f/3 beam with ADC correction over the 0.37--1.6\,$\mu$m wavelength range. Despite delivering an exceptional \'{e}tendue of approximately 350\,m$^2$\,deg$^2$ — roughly 50 times larger than the VLT focal plane and $\sim$10\% higher than Rubin/LSST — the optical design maintains excellent image quality, achieving an FWHM of 0.13 to 0.30\,arcsec at $z < 30^\circ$ over the full 2-degree field of view. Optical fibres are positioned in the MOS focal plane by a dedicated 32,000-unit robotic positioner.

The IFS optical path includes a 1-meter deformable mirror providing GLAO correction, conjugated to an altitude of 100\,m above the ground. The IFS optical train achieves excellent image quality with an RMS of 0.02--0.11\,arcsec depending on the subfield location within the patrol field. The GLAO system delivers 0.5\,arcsec FWHM image quality at 650\,nm with 40\% probability and 85\% sky coverage at the Galactic pole.

Advanced coatings are under development, particularly for the IFS path which foresees a large number of optical elements; the modelled telescope throughput is expected to reach average values of 79.8\% and 91.3\% for the IFS and MOS optical paths, respectively (see \citenum{WST2026-Coatings} for details). The telescope will be housed in a conventional alt-azimuth enclosure with an estimated diameter of 40\,m and height of 38\,m.

\subsection{Integral Field Spectrograph (IFS)}
The IFS provides spatially resolved spectroscopy over a $3\times3$\,arcmin$^2$ field of view (nine times larger than MUSE) with $0.25''\times0.25''$ spaxels, covering 370--930\,nm\footnotemark\ simultaneously at an average spectral resolution $R \sim 4800$. A mosaic mode extends the effective field to $9\times9$\,arcmin$^2$ by repositioning the science field within the 13\,arcmin diameter patrol area.

The instrument comprises 192 integral-field units, each consisting of an image slicer, an f/2.0 two-channel spectrograph, and two $6\mathrm{k}\times6\mathrm{k}$ CMOS detectors with 15\,$\mu$m pixels. The IFS station is located in the Coud\'{e} room and occupies a cylindrical envelope of 26.4\,m diameter and 6.8\,m height ($\sim$3700\,m$^3$).

In {\ttfamily pyetc\_wst}, the IFS is modelled with a blue (370--640\,nm) and a red (620--980\,nm) channel, with spectral sampling of 0.048\,nm\,pix$^{-1}$ and 0.064\,nm\,pix$^{-1}$, respectively, yielding $R \approx 4000$--$5000$ across the full wavelength range. A scientific CMOS detector with RON\,$\approx\,1.4$\,e$^-$ (combined dual-readout) and dark current of 2\,e$^-$\,hr$^{-1}$ is assumed.

\subsection{Low-Resolution Multi-Object Spectrograph (MOS-LR)}
The MOS-LR provides simultaneous spectroscopy for 30,000 targets across 370--930\,nm\footnotemark[\value{footnote}], partitioned into four contiguous spectral bands. It comprises 54 spectrograph units, each with four channels and $6\mathrm{k}\times6\mathrm{k}$ CMOS detectors (15\,$\mu$m pixels), located on the azimuth platform ($\sim$174\,m$^3$ total volume). The spectral resolving power ranges from $R \sim 3000$ in the first three bands to $R \sim 3500$ in the reddest band, which is optimised to improve sky-background subtraction at long wavelengths.
\footnotetext{The current ETC model extends to 980\,nm; throughput curves will be updated as the design is finalised.}

In {\ttfamily pyetc\_wst}, the MOS-LR is modelled with four channels (blue, green, yellow, red) covering 370--980\,nm, delivered to $1.03''$ fibres with readout noise of 1\,e$^-$ and dark current of 1\,e$^-$\,hr$^{-1}$.

\subsection{High-Resolution Multi-Object Spectrograph (MOS-HR)}
The MOS-HR provides $R \sim 40{,}000$ spectroscopy for 2,000 targets in four non-contiguous wavelength windows: 400.9--443.1, 452.2--499.8, 542.4--599.6, and 608.0--672.0\,nm, selected to cover key chemical-abundance diagnostics. Achieving this resolving power on a 12-m aperture requires splitting each $1''$ input fibre into multiple smaller fibres to reduce the effective slit width. Two architectures are currently under investigation: a 1-to-7 fibre-split design based on 8 spectrograph units ($16\mathrm{k}\times16\mathrm{k}$ CMOS, 10\,$\mu$m pixels), and a 1-to-19 split based on 16 smaller units ($10\mathrm{k}\times10\mathrm{k}$ CMOS, 9\,$\mu$m pixels). The final choice will depend on the trade-off between throughput, complexity, and cost.

\begin{table}[ht]
\caption{Summary of WST spectral channels as implemented in {\ttfamily pyetc\_wst}.
$\lambda_1$--$\lambda_2$: wavelength range; $R$: resolving power at channel
centre; spaxel/fibre: angular element size; RON: single-read readout noise.}
\label{tab:instruments}
\begin{center}
\small
\begin{tabular}{llccccc}
\toprule
Mode & Channel & $\lambda_1$--$\lambda_2$ (nm) & $R$ & Spaxel/fibre & RON (e$^-$) & Dark (e$^-$/hr) \\
\midrule
IFS  & Blue   & 370--640   & 4200 & $0.25''$ & 1.4 & 2 \\
IFS  & Red    & 620--930   & 5000 & $0.25''$ & 1.4 & 2 \\
\midrule
MOS-LR & Blue   & 370--477   & 3030 & $1.03''$ & 1.0 & 1 \\
MOS-LR & Green  & 463--608   & 2960 & $1.03''$ & 1.0 & 1 \\
MOS-LR & Yellow & 592--771   & 2910 & $1.03''$ & 1.0 & 1 \\
MOS-LR & Red    & 749--930   & 3510 & $1.03''$ & 1.0 & 1 \\
\midrule
MOS-HR & Blue   & 400.9--443.1 & 43400 & $1.0''$ & 1.0 & 1 \\
MOS-HR & Green  & 452.2--499.8 & 41300 & $1.0''$ & 1.0 & 1 \\
MOS-HR & Yellow & 542.4--599.6 & 36900 & $1.0''$ & 1.0 & 1 \\
MOS-HR & Red    & 608.0--672.0 & 37000 & $1.0''$ & 1.0 & 1 \\
\bottomrule
\end{tabular}
\end{center}
\end{table}

\section{ETC COMPUTATIONAL MODEL}
\label{sec:model}

\subsection{Photon budget and throughput}
\label{sec:photon}

The ETC models the complete photon path from the astronomical source to the
detector.  For each wavelength element $\lambda$, the detected signal in
photo-electrons per spectral pixel is
\begin{equation}
S(\lambda) = F_\lambda(\lambda)\,
             \frac{\lambda}{hc}\,A_{\rm tel}\,
             \tau_{\rm atm}(\lambda)\,
             T_{\rm ins}(\lambda)\,
             f_{\rm fib}(\lambda)\,
             t_{\rm exp}
\label{eq:signal}
\end{equation}
where $F_\lambda$ is the source spectral flux density at the top of the
atmosphere, $A_{\rm tel} = \pi(D/2)^2$ is the unobscured collecting area for a
12\,m primary mirror, $\tau_{\rm atm}$ is the atmospheric transmission,
$T_{\rm ins}$ is the total instrument throughput (telescope optics, spectrograph,
and detector quantum efficiency), $f_{\rm fib}$ is the fibre injection fraction
(1 for IFS spaxels; wavelength-dependent for MOS point and extended sources), and
$t_{\rm exp} = N_{\rm DIT}\times t_{\rm DIT}$ is the total on-source integration
time.

The instrument throughput $T_{\rm ins}$ is provided as wavelength-dependent
tables (one per channel, Fig.~\ref{fig:throughput}) derived from the throughput
model delivered by the WST system engineering team (O.\ Bellido, March 2026). 

\subsection{Spatial morphologies}
\label{sec:psf}
Three spatial morphologies are available.

\textbf{Point source.}  The on-sky PSF is modelled as a Moffat
profile\,\citenum{Moffat69} with $\beta = 2.8$ and a total FWHM given by
quadrature addition of the telescope image quality ($\theta_{\rm tel}$), the
instrument image quality ($\theta_{\rm ins}$), and the
wavelength-dependent atmospheric seeing scaled to the target airmass.  For
MOS modes, the fraction of the source flux enclosed within the fibre aperture
is either computed analytically (surface-brightness sources) or by integration of the
PSF image (point/resolved sources).

\textbf{Extended source (surface brightness).}  The source is treated as
uniform across the resolution element; the fibre injection fraction is 1.

\textbf{Resolved S\'ersic profile.}  A 2-D S\'ersic image is generated,
convolved with the PSF, and the fraction within the aperture computed
numerically.  The effective radius $r_e$ and S\'ersic index $n$ are
user-settable parameters.

\subsection{Sky background modelling}
\label{sec:sky}

Two sky-background modes are available.

\textbf{Static mode.}  Pre-computed sky spectra tabulated for three sky
brightness levels (dark, grey, bright, corresponding to moon illumination
FLI\,$=\,0$, $0.5$, $1.0$) and four airmass values ($X\,=\,1.0$, $1.2$,
$1.5$, $2.0$) at three precipitable-water-vapour (PWV) values
(1.0, 3.5, 10.0\,mm) are stored internally.  This mode requires no external
connection and is suitable for rapid surveys.

\textbf{SkyCalc mode.}  A live query is issued to the ESO SkyCalc web
service\,\citenum{Noll12,Jones13} via the {\ttfamily skycalc\_cli}
package\,\citenum{skycalc_cli}.  This delivers self-consistent sky emission and
atmospheric transmission spectra for arbitrary observing conditions (airmass,
FLI, PWV, moon--target separation), enabling accurate predictions for
specific observing programmes.  The sky model returns separate emission
($\mathrm{flux}$) and transmission ($\mathrm{trans}$) spectra which are
subsequently convolved to the instrument spectral resolution and rebinned onto
the channel wavelength grid.

\begin{figure}[ht]
\begin{center}
\includegraphics[width=0.75\textwidth]{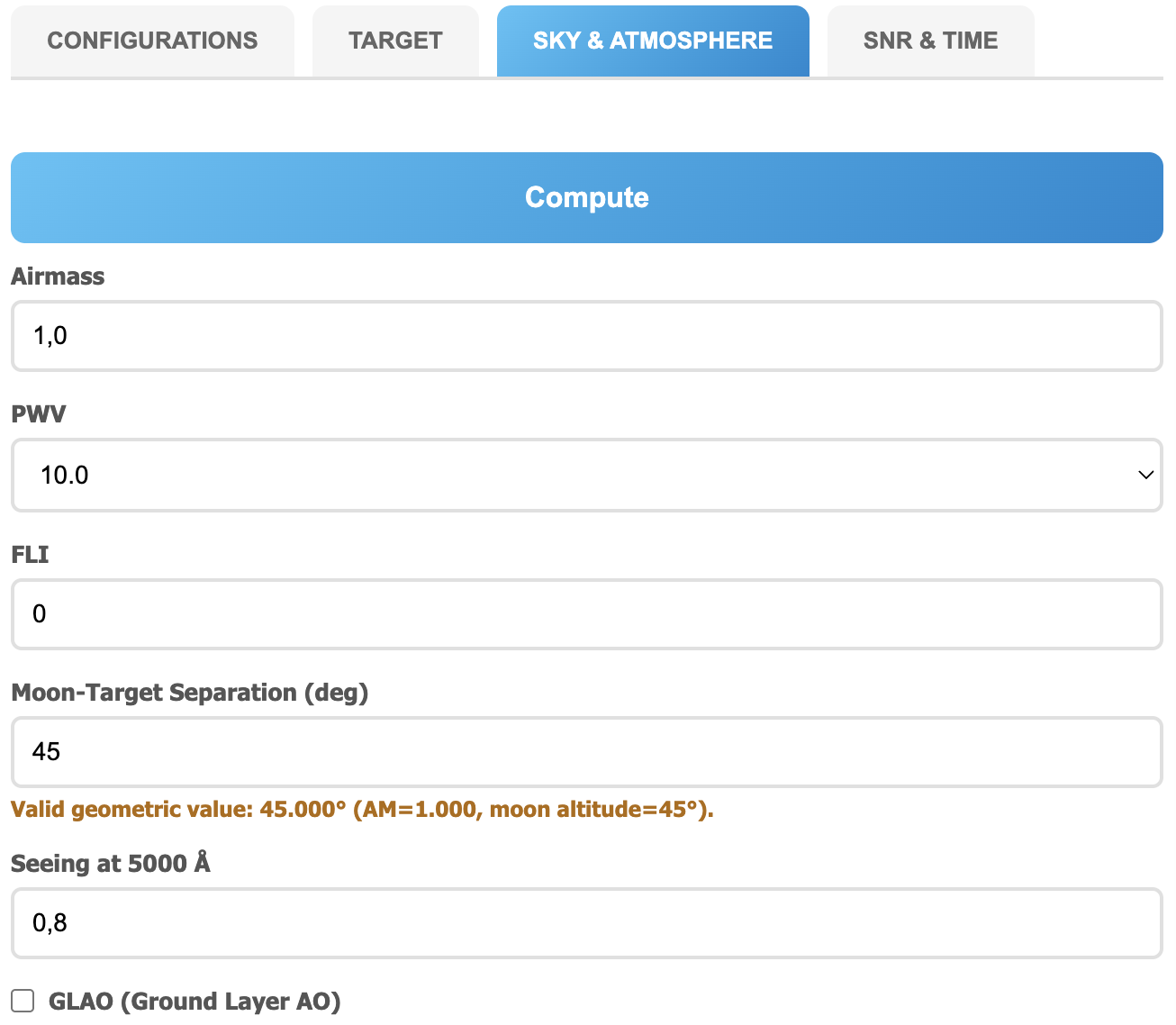}
\end{center}
\caption{\label{fig:web_sky}
The \textit{Sky \& Atmosphere} tab of the {\ttfamily pyetc\_wst} web interface
(see Section~\ref{sec:web}).
From top to bottom: airmass, precipitable water vapour (PWV, mm), lunar
fraction of lunar illumination (FLI, 0--1), moon--target angular separation
(degrees), and seeing FWHM at 5000\,\AA.
The interface validates the geometric consistency of the airmass and moon
altitude in real time (orange feedback line).
The \textit{GLAO} checkbox activates the Ground-Layer Adaptive Optics
correction for IFS observations, replacing the free-atmosphere seeing with
the GLAO-corrected PSF.}
\end{figure}

Figure~\ref{fig:web_sky} shows the \textit{Sky \& Atmosphere} tab of the web interface (Section~\ref{sec:web}).

\subsection{Source spectral models}
\label{sec:sed}

Five classes of spectral energy distribution (SED) are supported:

\begin{itemize}\setlength{\itemsep}{0pt}
  \item \textbf{Stellar templates}: a library of $>$400 spectra including
    MARCS models\,\citenum{Gustafsson08} (2900--8000\,K), Pickles stellar
    atlas\,\citenum{Pickles98}, Kurucz models\,\citenum{Kurucz93},
    PHOENIX models\,\citenum{Husser13} (2300--12000\,K,
    $\log g = 0$--6), Kinney galaxy templates\,\citenum{Kinney96}, and
    25 SWIRE galaxy/AGN empirical SEDs\,\citenum{swire_templates}.
  \item \textbf{Blackbody}: Planck function for a user-specified temperature.
  \item \textbf{Power law}: $F_\lambda \propto \lambda^{\alpha}$, with
    user-specified index $\alpha$.
  \item \textbf{Emission line}: a Gaussian profile at a specified wavelength and
    line width, useful for emission-line galaxy studies.
  \item \textbf{User upload}: an ASCII or FITS table of wavelength and flux
    density; optional comment headers select wavelength units (\AA\ or nm) and
    flux units (erg\,cm$^{-2}$\,s$^{-1}$\,\AA$^{-1}$ or photon\,cm$^{-2}$\,s$^{-1}$\,\AA$^{-1}$).
\end{itemize}
All templates support an arbitrary spectroscopic redshift $z$.  The source
magnitude is specified in the Vega or AB system through a library of 22
photometric bands: twelve Vega-system filters ($UBVRIJHK$, Gaia $G$,
$G_{\rm BP}$, $G_{\rm RP}$, $G_{\rm RVS}$) and ten AB-system filters
(SDSS $ugriz$ and LSST $ugriz$); if no magnitude is specified, the template
flux is used as-is.

\begin{figure}[ht]
\begin{center}
\includegraphics[width=0.75\textwidth]{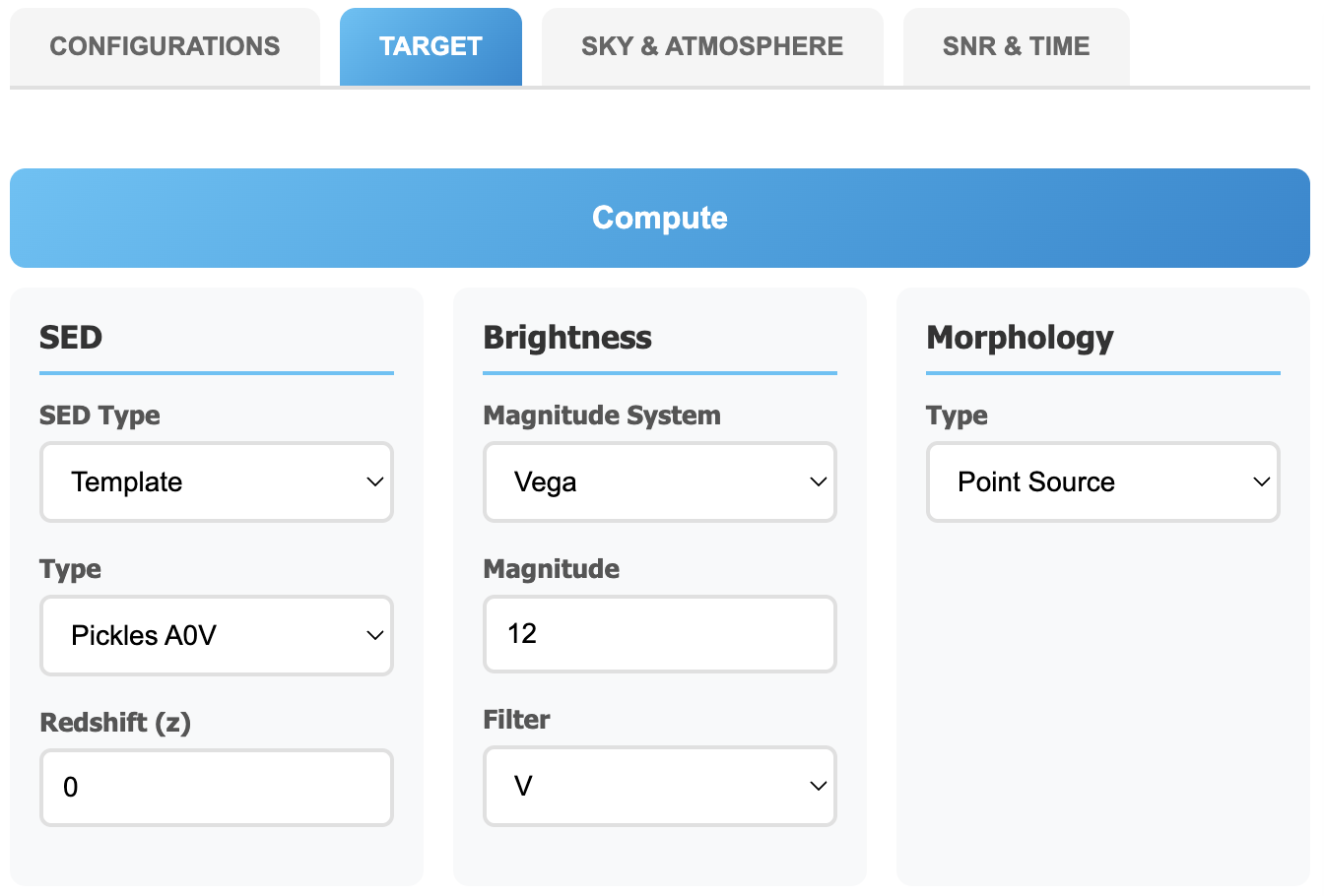}
\end{center}
\caption{\label{fig:web_target}
The \textit{Target} tab of the {\ttfamily pyetc\_wst} web interface.
\textit{Left}: spectral energy distribution selector (SED type, template, redshift).
\textit{Centre}: brightness panel (magnitude system, magnitude, and normalisation filter).
\textit{Right}: spatial morphology selector.}
\end{figure}

Figure~\ref{fig:web_target} shows the \textit{Target} tab of the web interface (Section~\ref{sec:web})

\subsection{Fibre Injection Fraction}
\label{sec:fib}

The ETC accounts for the efficiency with which source light is coupled into the
spectrograph optical fibre, known as the fibre injection fraction.  This is a
crucial factor in determining the total throughput of the MOS modes; for the IFS,
each $0.25''\times0.25''$ spaxel collects all light within its area, so the
injection fraction is unity by construction.

The on-sky source image is modelled as a two-dimensional Moffat
profile\,\citenum{Moffat69}, which reproduces the extended wings of
atmospheric seeing.  The fibre entrance is represented as a circular aperture,
and the injection fraction is computed by integrating the normalised PSF over
the fibre area.

The ETC also accounts for positional offsets between the PSF centroid and the
fibre centre — parameterised as the \textit{Object–fibre displacement} — which
can arise from telescope guiding errors or imperfect target acquisition.
Such misalignment reduces the injection fraction and can substantially degrade
the effective throughput for point sources.

\subsection{Noise model and SNR}
\label{sec:noise}

The total noise variance per co-added spectral pixel (coadded over
$N_{\rm sp}$ spectral pixels and $N_{xy}$ spatial pixels) is
\begin{equation}
\sigma^2 = N_{\rm DIT}
  \bigl[S_{\rm src} + S_{\rm sky} + N_{\rm pix}\,(d\,t_{\rm DIT} + \sigma_{\rm RON}^2)\bigr]
\label{eq:noise}
\end{equation}
where $S_{\rm src}$ and $S_{\rm sky}$ are the source and sky counts coadded over
$N_{\rm pix} = N_{\rm sp}\times N_{xy}$ pixels per DIT,
$d$ is the dark-current rate, and $\sigma_{\rm RON}$ is the per-pixel read-out
noise.  The SNR is $S_{\rm src}\,N_{\rm DIT}/\sigma$.

The ETC outputs the full noise decomposition (fraction of noise variance from
source, sky, RON, and dark current) as a function of wavelength, facilitating
regime identification and aiding detector trade-off studies.

Four computation modes are provided: (i)~compute SNR for fixed DIT and NDIT;
(ii)~compute the required NDIT for a target SNR and a fixed DIT; (iii)~compute
the required DIT for a target SNR and a fixed NDIT; (iv)~find the optimal
DIT--NDIT combination for a target SNR, minimising total readout noise while
satisfying user-defined DIT constraints.

\begin{figure}[ht]
\begin{center}
\includegraphics[width=0.75\textwidth]{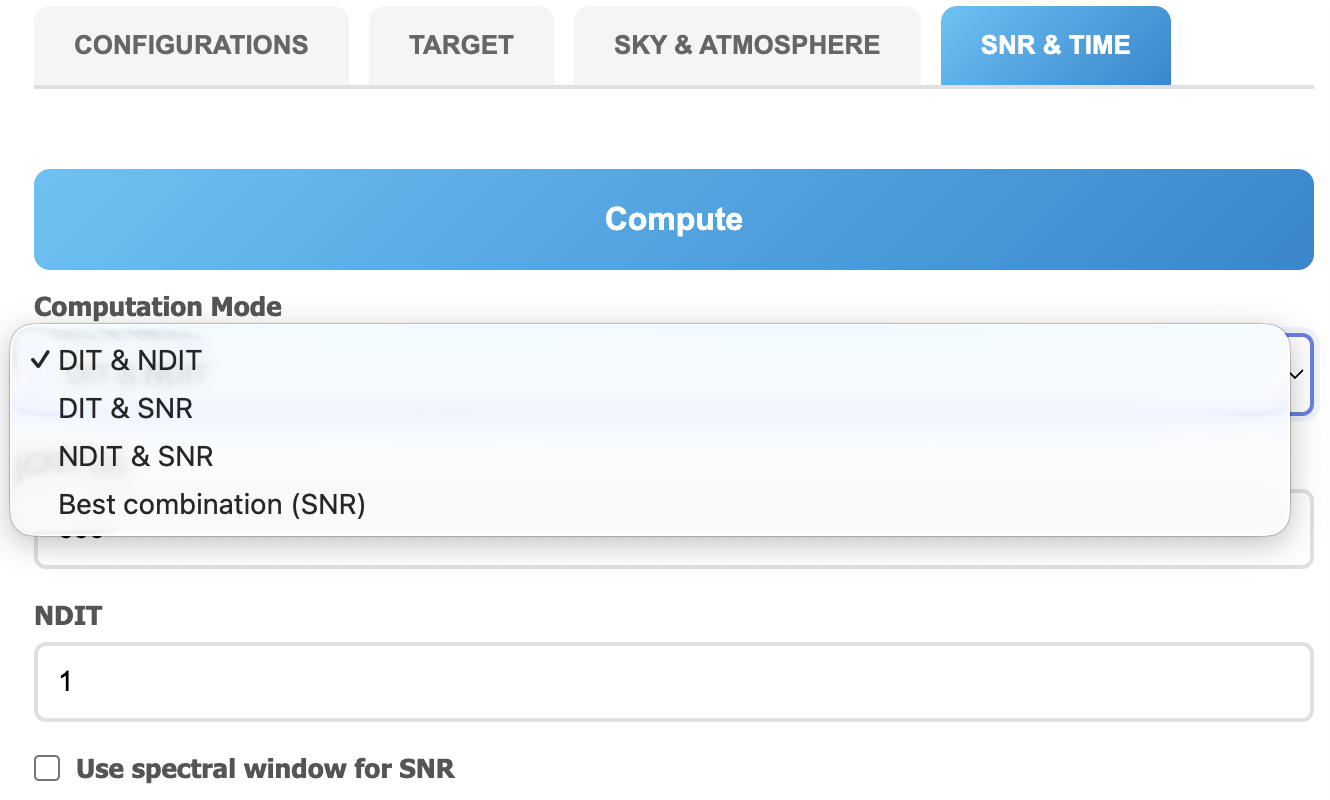}
\end{center}
\caption{\label{fig:web_comp}
The \textit{SNR \& Time} tab of the {\ttfamily pyetc\_wst} web interface
(see Section~\ref{sec:web}).  The drop-down menu exposes the four computation
modes: \textit{DIT \& NDIT} (compute SNR for fixed exposure parameters),
\textit{DIT \& SNR} (find NDIT for a target SNR),
\textit{NDIT \& SNR} (find DIT for a target SNR), and
\textit{Best combination} (optimise DIT--NDIT jointly).
The \textit{Use spectral window for SNR} checkbox activates averaging of the
SNR over a user-specified wavelength interval.}
\end{figure}

Figure~\ref{fig:web_comp} illustrates the \textit{SNR \& Time} tab, from the web interface (Section~\ref{sec:web}), which
exposes the four computation modes described in this section.

\subsection{1D Raw spectrum}

Beyond SNR predictions, the ETC can generate a Monte Carlo realisation of the
observed 1D spectrum for each channel.  For each wavelength element, source and
sky photon counts are drawn from Poisson statistics and combined with Gaussian
read-out noise; the result is a simulated extracted spectrum with its associated
noise as a function of wavelength.  The procedure is applied independently to
each selected channel, making it suitable for pipeline testing, detector
trade-off studies, and science-case visualisation.

\begin{figure}[ht]
\begin{center}
\includegraphics[width=0.85\textwidth]{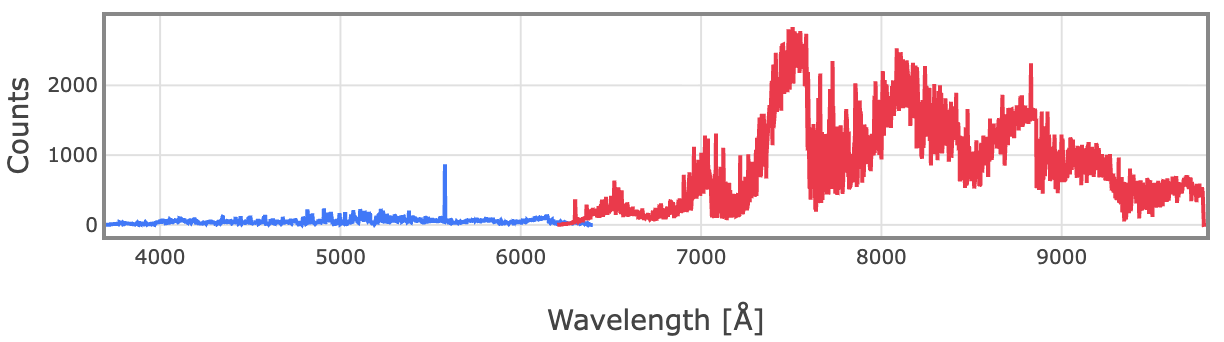}
\end{center}
\caption{\label{fig:sim_spec}
Simulated 1D raw spectrum for a MARCS cool-giant template, as recorded by the IFS blue and red channels. $T_{\rm eff}=3000$\,K, $\log g=3.0$
point source at $V=20$ (Vega), single exposure of 600\,s, dark sky
(FLI\,$=\,0$), airmass 1.0, seeing $0.8''$, PWV\,$=\,10$\,mm.
\textit{Blue}: IFS blue channel.
\textit{Red}: IFS red channel.
Counts are summed over the $1\times1$ spaxel extraction aperture with no
spectral co-adding.}
\end{figure}

Figure~\ref{fig:sim_spec} shows a simulated 1D raw spectrum example as displayed by
the {\ttfamily pyetc\_wst} web interface (Section~\ref{sec:web}).

\section{SOFTWARE ARCHITECTURE}
\label{sec:software}

\subsection{Python library: \texttt{pyetc\_wst}}
\label{sec:library}

The ETC is packaged as the {\ttfamily pyetc\_wst} Python
library\,\citenum{pyetc_wst} (current release 1.4, May 2026).  It is
distributed via GitHub and can be installed with {\ttfamily pip}.  The
dependency chain is minimal: NumPy\,\citenum{numpy}, SciPy\,\citenum{scipy},
Astropy\,\citenum{astropy}, MPDAF\,\citenum{Bacon16_mpdaf}, and
{\ttfamily skycalc\_cli}\,\citenum{skycalc_cli}.

The package is structured as follows.

\begin{itemize}\setlength{\itemsep}{0pt}
  \item {\ttfamily etc.py}: The {\ttfamily ETC} base class implementing the
    full computational pipeline: photon-budget evaluation, sky interpolation,
    PSF and fibre injection modelling, noise calculation, and all four
    computation modes.  Utility functions for spectrum operations, PSF image
    generation, and noise visualisation are also provided.
  \item {\ttfamily wst.py}: The {\ttfamily WST} class, subclassing {\ttfamily ETC},
    that loads all channel configurations, throughput tables, and pre-computed
    sky spectra.  It is the single access point for users and holds all version
    and changelog metadata.
  \item {\ttfamily specalib.py}: Photometric system definitions, SED template
    loading (with a class-level cache), filter profile management, and
    magnitude normalisation utilities.
\end{itemize}

A typical session requires a single call to {\ttfamily build\_obs\_full()} to
validate and expand a user-supplied parameter dictionary, followed by one of
{\ttfamily snr\_from\_source()}, {\ttfamily snr\_at\_wave()}, or
{\ttfamily time\_from\_source()}.  The {\ttfamily WST} object is designed to be
instantiated once and reused across many calculations, keeping the loaded
throughput curves and templates in memory.

\subsection{Web interface}
\label{sec:web}

A Flask-based web application ({\ttfamily pyetc\_web}) provides an interactive
browser interface.  Users configure the instrument mode, observing conditions,
and target parameters through a structured form.  Results are returned as
interactive Plotly plots (SNR spectrum, noise components, throughput curves) and
a numerical summary panel with per-channel output.  A save/load mechanism allows
JSON configuration files to be exported for reproducible calculations or used as
input to the programmatic interface. Examples of the interface panels are presented in Section~\ref{sec:model}, while examples of representative ETC output for selected science cases are shown in Section~\ref{sec:results}.

\subsection{REST API and command-line interface}
\label{sec:api}

A JSON-based REST API ({\ttfamily POST /api/compute}) exposes the full
computational functionality of the ETC.  All target, instrument, and observing
parameters accepted by {\ttfamily build\_obs\_full()} are passed as JSON fields;
the response contains all spectral outputs serialised as wavelength--value arrays
together with a scalar SNR result, noise fractions, and throughput data.

The companion script {\ttfamily wst\_cli.py} wraps the API for command-line use:
\begin{verbatim}
python wst_cli.py config.json -i 4 -o results.json
\end{verbatim}
where {\ttfamily config.json} is a parameter file (exportable from the web
interface).  Optional flags select JSON indentation ({\ttfamily -i}), output
redirection ({\ttfamily -o}), array collapse for readability ({\ttfamily -c}),
spectrum file upload ({\ttfamily -f}), and server address ({\ttfamily -s}),
making the CLI suitable for scripted survey-simulation pipelines.

\section{WEB PAGE INPUT AND RESULTS EXAMPLES}
\label{sec:results}
\begin{figure}[ht]
\begin{center}
\includegraphics[width=0.75\textwidth]{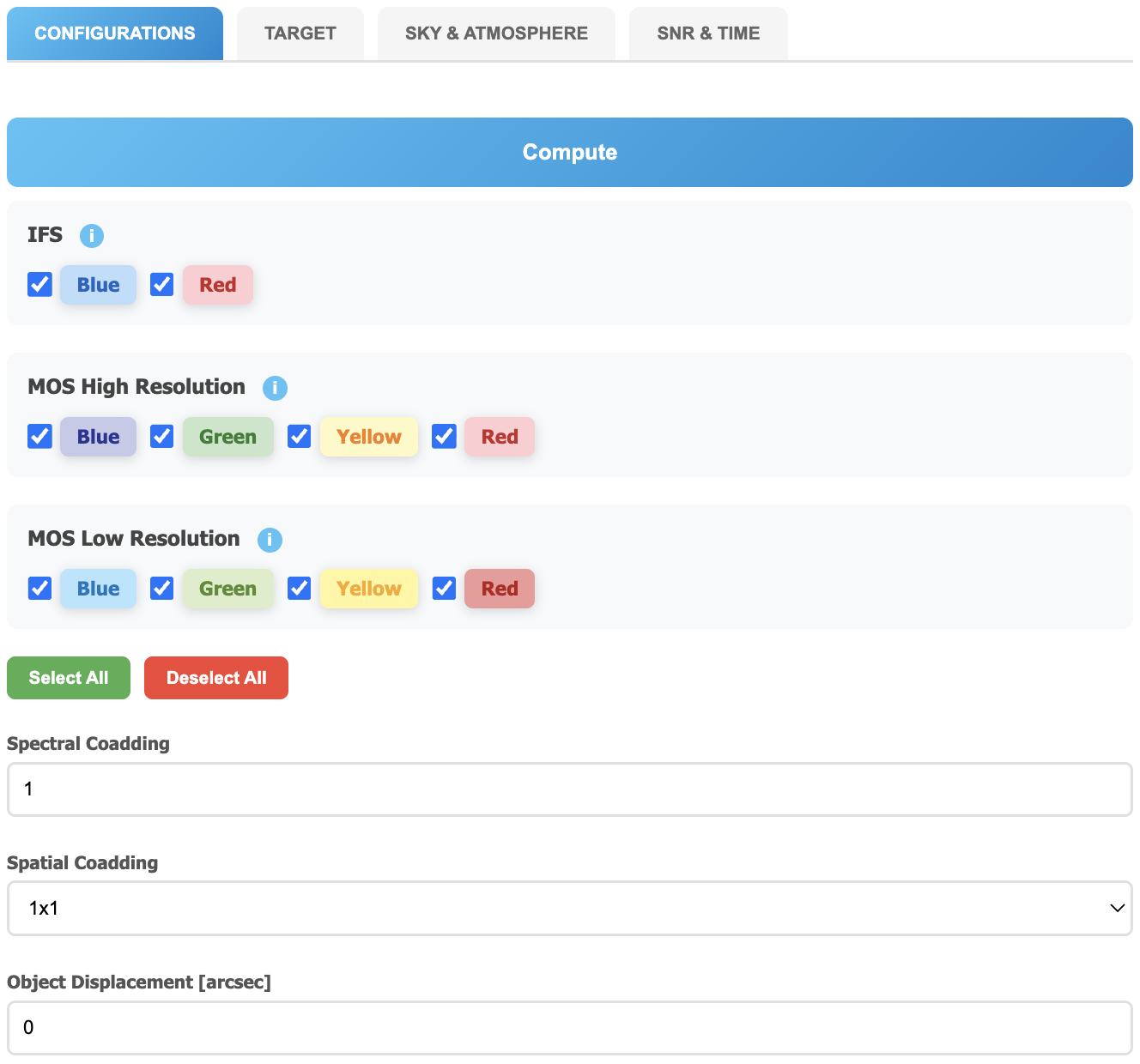}
\end{center}
\caption{\label{fig:web_conf}
The \textit{Configurations} tab of the {\ttfamily pyetc\_wst} web interface.  Instrument channels are selected individually
via colour-coded checkboxes grouped by mode (IFS, MOS-HR, MOS-LR), with
\textit{Select All} and \textit{Deselect All} shortcuts.  Below, the
spectral co-adding factor, spatial co-adding aperture ($N\times N$ spaxels
for IFS), and object--fibre displacement
(arcsec, for MOS) are set. All selected channels are computed simultaneously in a
single API call.}
\end{figure}

The ETC frontend application is an interactive browser interface which presents the user with four tabs, by which they can select the instrument configuration, the properties of the target, the sky conditions, and, finally, the computation mode.
The instrument configuration can set each possible instrument and channel combination, with related parameters (e.g. coadding), as shown in Figure~\ref{fig:web_conf}.

The other three main windows (\textit{Target}, \textit{Sky \& Atmosphere} and \textit{SNR \& Time}) are shown and discussed in Section~\ref{sec:model}.

\subsection{Instrument throughput}
\label{sec:throughput}

\begin{figure}[ht]
\begin{center}
\includegraphics[width=0.9\textwidth]{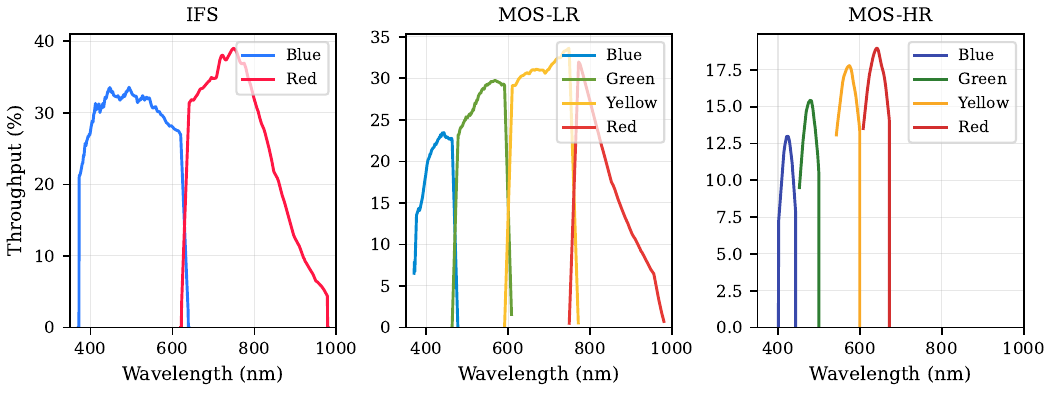}
\end{center}
\caption{\label{fig:throughput}
Instrument throughput (no atmosphere) for all WST spectral channels as a
function of wavelength (preliminary model, subject to revision as the design
matures). \textit{Left}: IFS blue and red channels.
\textit{Centre}: MOS-LR four bands.  \textit{Right}: MOS-HR four bands.}
\end{figure}

Figure~\ref{fig:throughput} shows the total instrument throughput (telescope,
spectrograph, detector; no atmosphere) for all ten WST channels.  The IFS red
channel peaks at $\approx\!39\%$ near 750\,nm, while the blue channel reaches
$\approx\!34\%$ near 490\,nm.  MOS-LR throughputs range from $\sim\!24\%$ in
the blue band to $\sim\!34\%$ in the yellow band, with the green band reaching
$\sim\!30\%$.  MOS-HR throughputs span $\sim\!13$--$19\%$, consistent with the
additional high-dispersion optical elements and the smaller fibre etendue.
These curves represent a preliminary throughput model delivered by the WST system
engineering team; they will be updated as the design matures, and all performance numbers reported in this paper should be interpreted accordingly.

\subsection{Signal-to-noise ratio spectra}
\label{sec:snr_spectra}

\begin{figure}[ht]
\begin{center}
\includegraphics[width=0.85\textwidth]{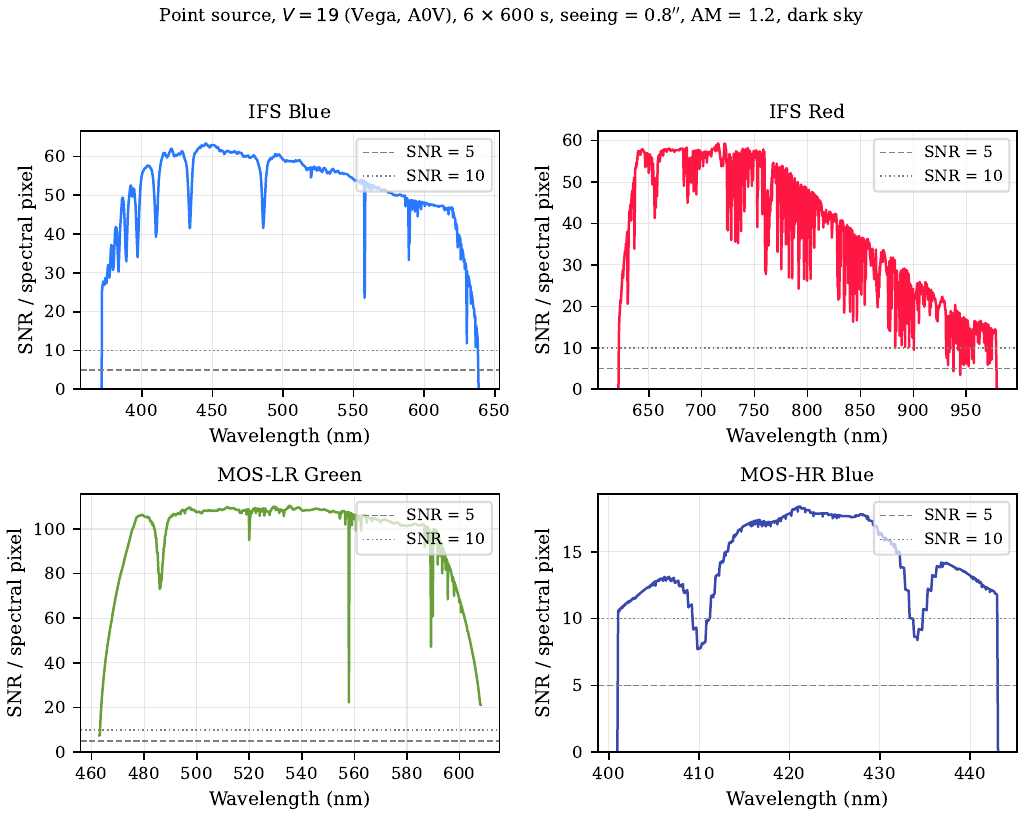}
\end{center}
\caption{\label{fig:snr}
SNR per spectral pixel (rebinned by 5\,\AA\ for display) for a $V=19$ A0V
star, $6\times 600$\,s, seeing\,$=0.8''$, AM\,$=1.2$, dark sky,
PWV\,$=3.5$\,mm.  \textit{Top left}: IFS Blue.
\textit{Top right}: IFS Red.
\textit{Bottom left}: MOS-LR Green.
\textit{Bottom right}: MOS-HR Blue.
Horizontal dashed and dotted lines mark SNR\,$=$\,5 and 10, respectively.}
\end{figure}

Figure~\ref{fig:snr} shows example SNR spectra for a $V=19$ (Vega) A0V spectral-type star 
observed for $6\times600$\,s (1\,hr) under typical conditions (seeing
$0.8''$, airmass 1.2, dark sky, PWV\,=\,3.5\,mm). These IFS calculations use
natural seeing without GLAO correction; GLAO is applied only for the
surface-brightness limiting-magnitude computation in Section~\ref{sec:limmag},
where the improved image quality (see Section~\ref{sec:telescope}) reduces the
sky background per spaxel.  In the IFS channels the
peak SNR reaches $\sim\!60$\,pix$^{-1}$ (blue) and $\sim\!55$\,pix$^{-1}$
(red), well above 10 across the full spectral range. For MOS-LR, the green
channel exceeds SNR\,$=\,100$ at peak, while for MOS-HR the blue channel
gives SNR\,$\approx\!10$--$18$ for the same source magnitude, reflecting the
higher dispersion of the high-resolution mode. The absorption features visible in the spectra arise from the stellar template and atmospheric absorption.

\subsection{Limiting magnitudes}
\label{sec:limmag}

\begin{figure}[ht]
\begin{center}
\includegraphics[width=0.9\textwidth]{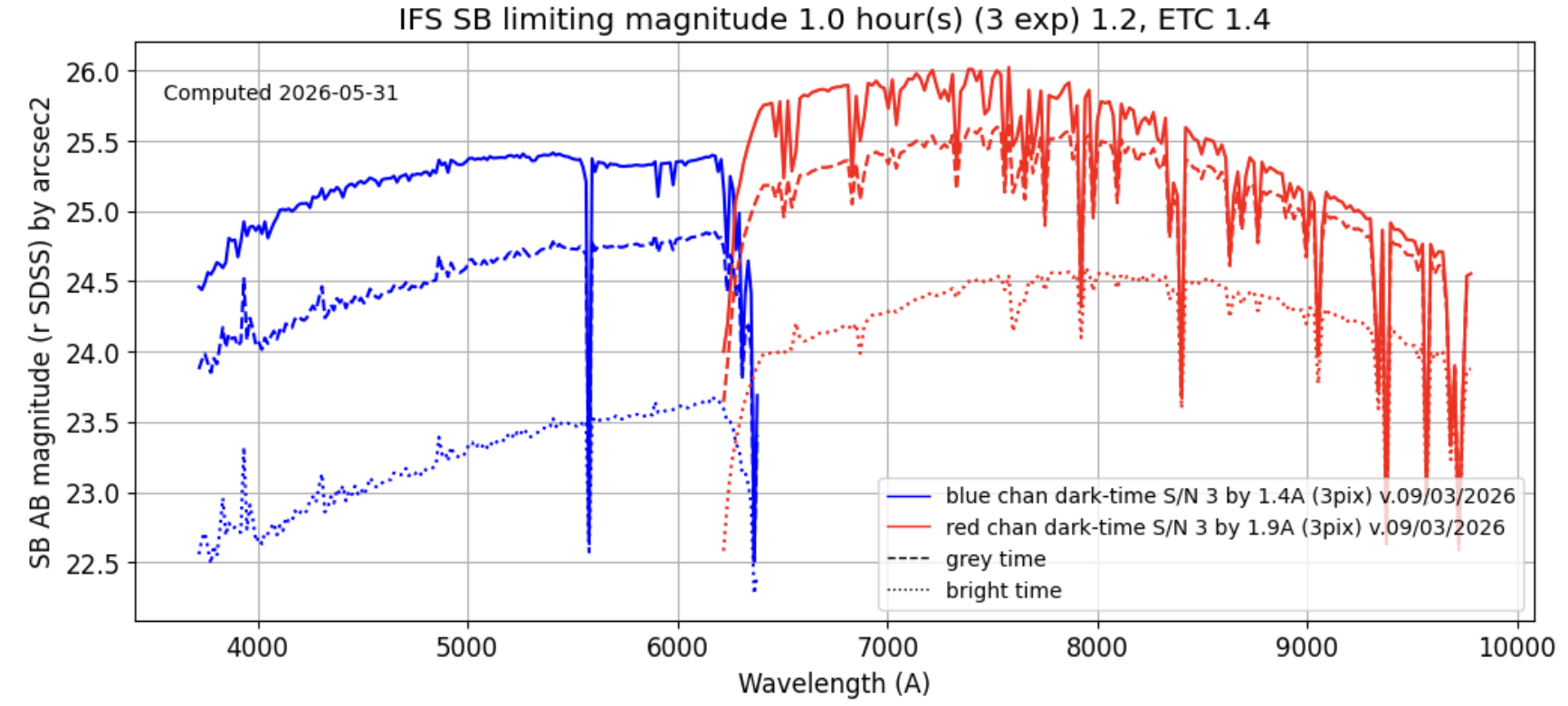}
\end{center}
\caption{\label{fig:limmag_ifs}
IFS surface-brightness limiting magnitude (SB AB, $r$-SDSS per arcsec$^2$;
SNR\,$=\,3$ per resolution element of 3 spectral pixels, $3\times1200$\,s,
GLAO, AM\,$=1.2$) as a function of wavelength for dark (solid), grey (dashed),
and bright (dotted) sky.  Blue and red channels are shown separately.}
\end{figure}

\begin{figure}[ht]
\begin{center}
\includegraphics[width=0.9\textwidth]{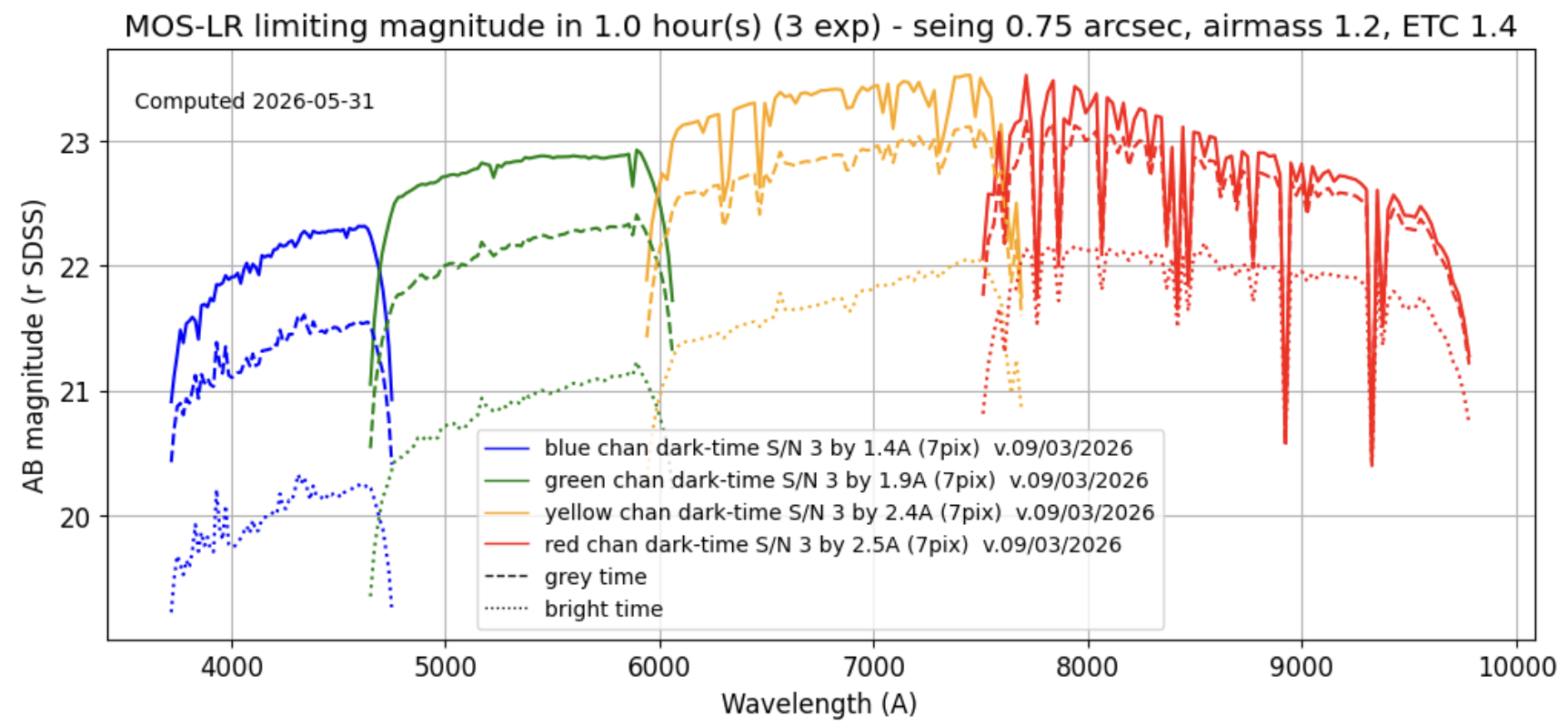}
\end{center}
\caption{\label{fig:limmag_mos}
MOS-LR limiting magnitude (AB, $r$-SDSS; SNR\,$=\,3$ per resolution element,
$3\times1200$\,s, seeing $0.75''$, AM\,$=1.2$) as a function of wavelength for
dark (solid), grey (dashed), and bright (dotted) sky.  The four channels
(blue, green, yellow, red) are shown separately.}\end{figure}

Figure~\ref{fig:limmag_ifs} shows the IFS surface-brightness (SB) limiting
magnitude as a function of wavelength, computed with GLAO correction for
$3\times1200$\,s at airmass 1.2.  Figure~\ref{fig:limmag_mos} shows the
MOS-LR point-source limiting magnitude under the same exposure conditions with
natural seeing of $0.75''$, for a flat-spectrum source normalised to $r$-SDSS.

Under dark-sky conditions, the IFS blue channel reaches
$\mu_r\approx25.4$\,AB\,arcsec$^{-2}$ near 500\,nm (SNR\,$=\,3$ per
resolution element of 1.4\,\AA\ co-added over 3 spectral pixels), with the
sensitivity decreasing towards the UV edge.  The IFS red channel spans
$\mu_r\approx25.5$--$26.0$\,AB\,arcsec$^{-2}$ across 650--850\,nm.
The dark-to-bright sky penalty reaches $\sim2$\,mag in the most
sky-background-limited spectral regions.

For MOS-LR, the four channels span $r\approx22$--$23.4$\,AB (dark sky,
SNR\,$=\,3$ per resolution element, 7\,pix), with the yellow channel reaching
the highest limiting magnitude owing to its favourable sky-background level.
The bright-sky penalty is most severe in the blue channel, exceeding
$1.5$\,mag at long wavelengths.

\subsection{Noise decomposition}
\label{sec:noise_decomp}

\begin{figure}[ht]
\begin{center}
\includegraphics[width=0.62\textwidth]{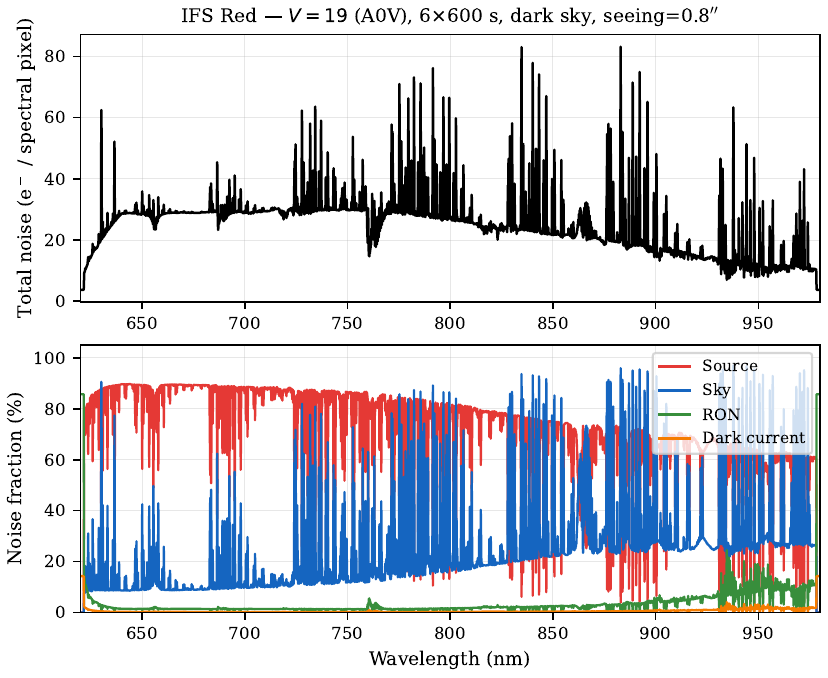}
\end{center}
\caption{\label{fig:noise}
Noise decomposition for IFS Red, $V=19$ A0V source, $6\times600$\,s, dark sky.
\textit{Top}: total noise in e$^-$ per spectral pixel (rebinned by 15\,\AA).
\textit{Bottom}: fractional noise contributions by source, sky, read-out noise
(RON), and dark current.}
\end{figure}

Figure~\ref{fig:noise} shows the noise decomposition for IFS Red at $V=19$.
At wavelengths below $\approx\!850$\,nm, source photon noise dominates (fraction
$\sim\!80\%$), confirming that a $V=19$ source is detected well above the
sky-background-limited regime. Between 860 and 980\,nm, sky emission bands
become increasingly important, driving the sky-noise fraction up to $\sim\!40\%$.
RON and dark current contribute less than 5\% throughout, confirming that the
assumed low-noise detector performance is not a limiting factor at this
magnitude.

\section{CONCLUSIONS}
\label{sec:conclusions}

We have described {\ttfamily pyetc\_wst}, the Exposure Time Calculator for the
WST instrument suite.  The tool delivers full end-to-end simulation of all three
spectrograph modes — IFS, MOS-LR, and MOS-HR — covering the complete 370--930\,nm
optical range.  Key features include:

\begin{itemize}\setlength{\itemsep}{0pt}
  \item Integration with the ESO SkyCalc service for accurate, on-the-fly sky
    background modelling;
  \item A rich library of target templates and SEDs and spatial morphologies;
  \item Four flexible computation modes for exposure-time optimisation;
  \item A web interface, a REST API, and a command-line tool enabling both
    interactive and batch usage;
  \item A fully open Python library installable via {\ttfamily pip}, promoting
    reproducibility and community contributions.
\end{itemize}

Performance examples confirm that WST reaches a surface-brightness limit of
$\mu_r\approx25.4$\,AB\,arcsec$^{-2}$ (IFS blue, dark sky, SNR\,$=\,3$
per resolution element, 1\,hr with GLAO at AM\,$=1.2$) and
$\mu_r\approx25.5$--$26.0$\,AB\,arcsec$^{-2}$ in the IFS red channel.
For MOS-LR, point-source dark-sky limits span $r\approx22$--$23.4$\,AB
across the four bands under the same conditions, demonstrating the survey
depth achievable with WST for wide-field spectroscopic programmes.

The tool is publicly available as a Python library and web service, and constitutes a complete, validated framework for WST survey design and science
case assessment.  The throughput model, sky-background treatment, and noise model have been verified against first-principles expectations and are consistent with the current WST system-engineering baseline.  As the WST
design matures, the ETC will be updated with revised throughput curves and
functionalities.

\acknowledgments

This project has been funded by the European Union’s Horizon Europe research and innovation programme under grant agreement No. 101183153.
The {\ttfamily pyetc\_wst} package builds upon the
original {\ttfamily pyetc} library developed by R.\ Bacon. We thank the WST System Engineer for providing the throughput model, and the
ESO SkyCalc team for maintaining the sky background web service.

\clearpage
\bibliography{report}
\bibliographystyle{spiebib}

\end{document}